\documentclass{article}

\usepackage{epsfig,sprocl}

\newcommand{\bear}{\begin{eqnarray}}
\newcommand{\ear}{\end{eqnarray}}

\newcommand{\gsim}{\mathrel{\vcenter
    {\hbox{$>$}\nointerlineskip\hbox{$\sim$}}}}

\begin{document}

\title{SOFT HIGH ENERGY SCATTERING IN NONPERTURBATIVE QCD}

\author{E. R. Berger}

\address{LAPTH, Universite Paris-Sud, Batiment 211,
\\ F-91405 Orsay Cedex, France\\E-mail: berger@th.u-psud.fr} 


\maketitle\abstracts{ In this report diffractive 
high energy reactions are discussed in a
functional integral approach where hadronic amplitudes are calculated
from vacuum expectation values of lightlike Wegner-Wilson loops.
In the first part we 
calculate elastic differential cross sections for  high energy and small
momentum transfer elastic proton-proton (pp) scattering which are in 
reasonable agreement with the experimental data.
In the second part we consider exclusive $\pi^0$ 
production in ep-scattering.
At high energies photon 
and odderon exchange contribute to this reaction. 
We show that odderon
exchange leads to a much larger inelastic than elastic $\pi^0$ production
cross section, dominating the $\gamma$ contribution by orders of
magnitude. Observing our process at HERA would establish the soft odderon.}

\section{Elastic pp scattering}

We will consider here some results of \cite{naber} for elastic differential 
cross sections $d \sigma / d t$ in elastic pp(p$\bar{{\rm{p}}}$)
and of \cite{etal} for exclusive $\pi^0$ production in ep scattering, both at 
high center of mass (cm) energies $\sqrt{s}\gsim
20$ GeV and low momentum transfer squared
$|t|{\mathrel{\vcenter
    {\hbox{$<$}\nointerlineskip\hbox{$\sim$}}}}
    \rm O (1\,{\rm GeV}^2)$ 
calculated within a functional approach \cite{na91,dfk}.

We start with elastic pp scattering. For 
large energies the amplitude $T_{\rm{pp}} (T_{{\rm{p}}\bar{{\rm{p}}}})$
is known to be dominated by $C=P=+1$ (pomern ${\cal{P}}$) exchange,
$C=P=-1$ (odderon ${\cal{O}}$) exchange\footnote{
The soft (nonperturbative) odderon is introduced in elastic 
hadron-hadron scattering as the $C=P=-1$ partner of the pomeron
\cite{nico1}.} is small or absent. In our approach, where the
p is treated as a wave packet of its constituent quarks, this is achieved
by considering the proton to be a quark-diquark (q-qq) system \cite{odd}, what
we will do in the following. Then the result for  $T_{\rm{pp}}$ is: 
\begin{eqnarray}
  &&T_{\rm{pp}} = (-2is) \int d^2b_T 
    {\exp}(i\vec{q}_T \vec{b}_T) \cdot
    \nonumber\\
  && \hphantom{T_{fi}}
   \int d^2x_T \, d^2y_T \, 
   | \Psi^{\rm{p}}(x_T) |^2 | \Psi^{\rm{p}}(y_T) |^2
   \tilde{J}(\vec{x}_T,\vec{y}_T,\vec{b}_T); 
  \label{ampl}
\end{eqnarray}
%
%
The scattering amplitude (\ref{ampl}) is obtained by first considering the
scattering of quarks (antiquarks) on a fixed gluon potential $G$.
Travelling through it the
quarks (q) and diquarks (qq) pick up non-abelian phase factors.
To ensure  gauge invariance the phase factors for
q and qq of the  same p
are joined at large positive and negative
times, yielding lightlike Wegner-Wilson loops $W_{\pm}$.
The parton scattering amplitude $\tilde{J}$, given by the functional integral
over $G$ (indicated with barackets $\langle \, \rangle_G$),
\bear
\tilde{J}(\vec{x}_T,\vec{y}_T,\vec{b}_T)=
    \Big\langle 
    W_+(\frac{1}{2} \vec{b}_T,{\vec x}_T)
    W_-(-\frac{1}{2} {\vec b}_T,{\vec y}_T)-1 
    \Big\rangle_G,
\label{jtilde}
\ear
%
%
is the central object in our approach. 
%
%
%
\begin{figure}[t]
  \unitlength1.0cm
  \begin{center}
    \begin{picture}(15.,9)

      \put(2.6,3){
        \epsfysize=7.7cm
        \epsffile{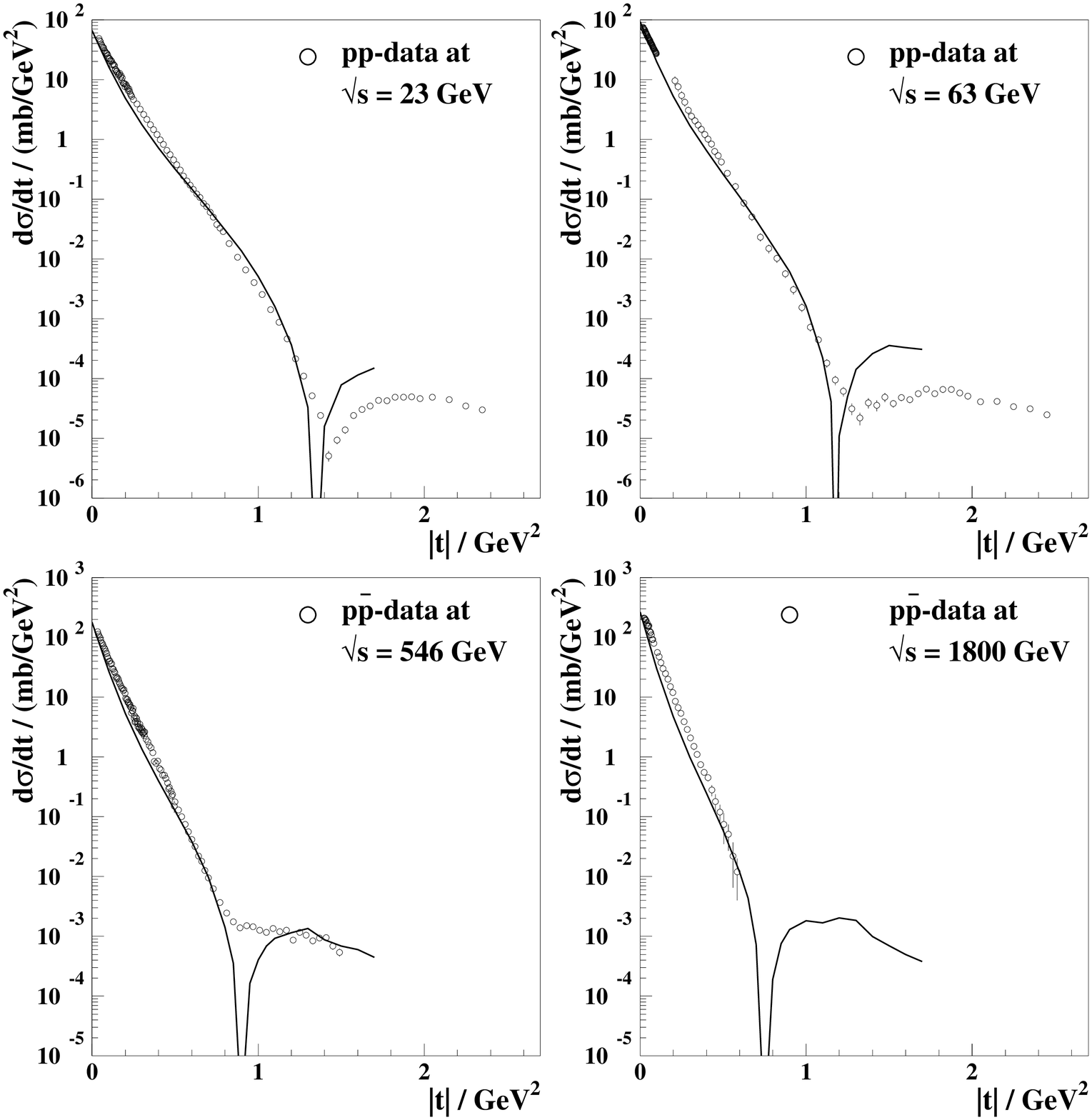}}

    \end{picture}
  \end{center}
\vspace*{-3.5cm}
\caption{ 
Differential elastic cross sections for
c.m. energies $\sqrt{s}=23,63,546$ and $1800${\rm{ GeV}}.}
  \label{probild}
\end{figure}
%
%
The transverse separation
between the centres of the loops (impact parameter) is 
given by $\vec{b}_T$. The
vectors $\vec{x}_T$ and $\vec{y}_T$ give the transvers extensions and
orientations of the loops. The resulting parton
scattering amplitude has to be integrated over $\vec{x}_T,\vec{y}_T$ 
with a measure given by
the q-qq densities, $\Psi^{\rm{p}} \sim exp(-\vec{x}^2/(4
S_{\rm{p}}^2))$. Here $S_{\rm p}$ is the p transvers extension parameter.
To calculate (\ref{ampl}) a
Fourier transform over 
$\vec{b}_T$ has to be done finally. 

The parton amplitude $\tilde{J}$ is calculated in nonperturbative QCD, 
using the minkowski version of the model of the stochastic vacuum (MSV). 
In terms of a matrix cumulant expansion the result is \cite{naber}:
\begin{eqnarray}
  &&\tilde{J}=
  \big[\frac{2}{3}
  \cos (\frac{1}{3}
  \chi  \,) +
  \frac{1}{3}
  \cos (\frac{2}{3}
  \chi  \,) - 1\big]  +
  i \big[ -\frac{2}{3}
  \sin (\frac{1}{3}
  \chi  \,)   \; +\frac{1}{3}
  \sin (\frac{2}{3}
  \chi  \,) \; \big] .
  \label{jstark}
\end{eqnarray}
The real function $\chi$ is calculated from an integral over the non-local gluon
condensate, for which the MSV makes an ansatz \cite{dfk,naber}. The parameters
that occur are the gluon condensate $G_2$, the correlation lenght $a$
and $\kappa$, a parameter measuring the non-abelian content of the
condensate. 
In this way high energy scattering is related to the QCD 
vacuum parameters.

Now $\chi$ is odd
under space inversion or charge conjugation, 
for example $\chi(-\vec{x}_T,\vec{y}_T,\vec{b}_T)=
-\chi(\vec{x}_T,\vec{y}_T,\vec{b}_T)$. For this reason $\tilde{J}$
contains $\cal{P}$ and $\cal{O}$ contributions. But since the q-qq densities in
(\ref{ampl}) are invariant under space inversion the $\cal{O}$ contributions
to $T_{\rm{pp}}$ average 
out consistent with experiment\footnote{A consequence is, 
that $T_{\rm{pp}}$ and $T_{\rm{p \bar{p}}}$ are equal in the diquark
limit.They contain no real part which is due to a second cumulant 
approximation \cite{naber}. As we will see, this results in
infinitly deep dips in $d \sigma / dt$}.

Leaving $G_2,\kappa , a$ and $S_{\rm p}$ in (\ref{ampl}) as free parameters 
we determine them in high energy scattering, using as input experimental 
data at $\sqrt{s}=23$ GeV. We obtain \cite{naber}:  
$G_2=(529 \, {\rm{MeV}})^4,\kappa=0.75, a=0.32 \, {\rm{fm}}$, 
consistent with existing lattice
measurements \cite{megg}, and a loop extension parameter $S_{\rm{p}}$ 
in the range of the electromagnetic proton radius, 
$S_{\rm{p}}=0.87$ fm.

The term of $\chi$ proportional to $\kappa$ results into string-string
interaction. This causes 
$\sigma_{tot}$ to increase with
increasing extension parameter \cite{dfk,naber}. Following
\cite{dfk} we introduce an energy dependence of (\ref{ampl}) by
making $S_p$ energy dependent, fixed by requiring that our 
model reproduces the pomeron part of $\sigma_{tot}$ in the DL 
parametrisation (The vacuum parameters should not depend on the energy).

In Fig. \ref{probild} we show our results for $d \sigma/dt$.
A major point is, that our results for $d\sigma/dt$
depend crucially on the string-string interaction,
$\kappa$$\neq$$0$,
what also implies in terms of the MSV a string tension
$\rho$$\neq$$0$ and so confinement.
A detailed discussion of this point can be 
found in \cite{naber}.
%
%
%
\begin{figure}[t]
  \unitlength1.0cm
  \begin{center}
    \begin{picture}(15.,4.0)

      \put(2,1.1){
        \epsfysize=4.0cm
        \epsffile{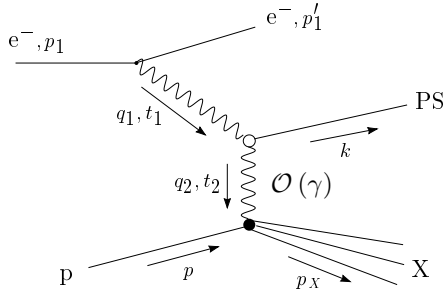}}

      \put(5.7,2.5){$\cal{O} \, (\gamma)$}

    \end{picture}
  \end{center}
  \vspace*{-1.3cm}
  \caption{Feynman diagrams for pseudoscalar meson production in $ep$
scattering at high energies with odderon (photon) exchange.}
\label{b1}
\end{figure}
%
%
%

\section{Exclusive $\pi^0$ production in ep scattering}

In this section we discuss 
exclusive $\pi^0$ production
in high energy ep scattering.
The $\pi^0$ is produced 
(see Fig. \ref{b1}) by
$\gamma \cal{O}$-fusion and by $\gamma \gamma$-fusion, but can not 
be produced by $\gamma \cal{P}$-fusion since the $\pi^0$ has positive 
$C$-parity. 

In Fig. \ref{b1} X stands for a proton or resonances or a sum over resonances.
Here we treat the very small $Q^2:= -q_1^2$ range 
where the ep cross section can be calculated
by folding the $\gamma$p cross section with the equivalent photon
spectrum of the electron. 

To calculate the $\cal{O}$ exchange 
we look at $\gamma p$-scattering in the c.m. system.
In the following we
discuss the cases, (i) that the proton, 
stays intact or (ii)
gets diffractively exited into the
resonances
$N(1520)$ with $J^{P}={\frac{3}{2}}^- $ 
and $N(1535)$ with $J^{P}={\frac{1}{2}}^- $,
described as exited quark-diquark systems. 

When considering unpolarised cross sections,
summed over both resonances in case (ii), 
the quark spin degree of freedom becomes
irrelevant and the calculation reduces to one where 
a spinless state stays intact  or is
exited to a 2P resonance. The amplitudes are again given by folding
$\tilde{J}$ with approperiate wave functions. 
For (ii) the helicity amplitudes are \cite{etal}
\begin{eqnarray}
 &&T(s_2,t_2)_{\lambda,\lambda_{\gamma}} = 
 - 2is_2 \int \, d^2 b_T \, e^{i\vec{q_2}_T \vec{b}_T}\,
 \hat{J}_{\lambda,\lambda_{\gamma}} (\vec b),
 \nonumber\\
 && \hat{J}(\vec{b})_{\lambda,\lambda_{\gamma}} = 
 \int \frac{d^2 x_T}{4\pi}
 \int \frac{d^2 y_T}{4\pi} 
\nonumber\\
 &&\hphantom{\hat{J}(\vec{b})}
\times
\sum_{f,h_1, h_2} 
 \Psi^{*\, \pi^0}_{f h_1 h_2}(\vec{x}_T)
 \Psi^{\gamma}_{\lambda_{\gamma},\,f h_1 h_2} (\vec{x}_T)
 \nonumber\\
 &&\hphantom{\hat{J}(\vec{b})}
 \times
 \Psi^{*\, {\rm 2P}}_{\lambda} (\vec{y}_T)
 \Psi^{p}(\vec{y}_T) \; \;
 \tilde{J}(\vec{x}_T, \vec{y}_T,\vec{b}_T).
\label{psamp}
\end{eqnarray}
Here $\lambda_{\gamma} (\lambda)$ is the helicity and 
$\Psi^{\gamma}(\Psi^{\rm 2P})$ the wave function of the photon (2P
state).
%
%
%
%
\begin{figure}[t]
  \unitlength1.0cm
  \begin{center}
    \begin{picture}(15.,4.0)
      \put(2.5,0.8){
        \epsfysize=5.5cm

        \epsffile{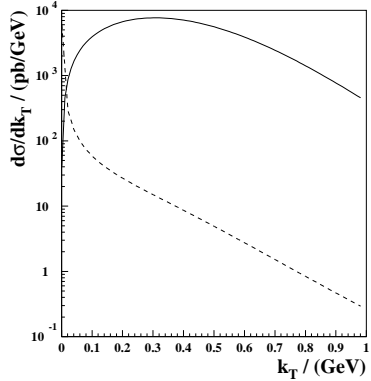}}
    \end{picture}
  \end{center}
  \vspace*{-1.1cm}

  \caption{The $k_T$ distribution in pion production
from the 2P resonance channels for 
odderon exchange (solid line)
compared to the
complete electromagnetic result (dashed line). Interference 
contributions are not taken into account.}
\label{k_Tmsv}
\end{figure}
%
%
%
Now, when the proton stays intact there occurs in (\ref{psamp})
the quark-diquark density
($\Psi_{\lambda}^{\rm 2P} \rightarrow \Psi^{\rm p}$). 
But, since this 
density is  symmetric under a parity transformation 
whereas the odderon coupling changes
sign, there is a cancelation when we integrate 
over all angles. On the other hand when the
proton gets exited to a negative parity state like the resonances
$N(1520)$ and $N(1535)$ there is no cancellation (the overlap is odd
under parity) and the
odderon couples to the nucleon without any restriction 
\cite{etal,donaru}. Thus in our model the odderon couples only if  
breakup of the proton occurs. 

The photoproduction cross section calculated from (\ref{psamp}) is 
independent of $s_2$ and so
the EPA conversion to electroproduction can be achieved by simply multiplying
it with a constant $c_{{\rm EPA}} = 0.0136$. 

In the following we consider only ep scattering.
An experimentally prefered observable is the $k_T$ spectrum,
the transverse momentum distribution of the $\pi^0$ with respect 
to the beam direction. 
This is displayed in Fig. \ref{k_Tmsv}. 
We also show there the 
distribution of the pion's $k_T$ for $\gamma$ exchange
summed over the elastic and 
all inelastic channels (with invariant mass $M_X \le 2$ GeV). 
As we can see $\gamma$ exchange is larger than $\cal{O}$ exchange only
for very small $k_T$. For $k_T \gsim 0.1$ GeV the $\cal{O}$ exchange
dominates by orders of magnitude.

\section{Conclusions}

The functional integral approach gives a reasonable description
of soft elastic pp scattering at high energies, relating it to
fundamental parameters of the QCD vacuum. The values determined from
high energy scattering compare well with lattice results. In $\pi^0$ production
for the p breakup into negative parity hadronic final states, 
a large odderon cross section arises which
should be observable at HERA. 
This would establish the soft odderon as an exchange-object
in high energy scattering on an equal footing with the soft pomeron.


\section*{References}
\begin{thebibliography}{99}

\bibitem{naber} E. R. Berger, O. Nachtmann,
  Eur. Phys. J. C7, 459 (1999)

\bibitem{etal}
  E. R. Berger, A. Donnachie, H. G. Dosch,  
  W. Kilian, O. Nachtmann, M. Rueter, 
  Eur. Phys. J. C9, 491 (1999);

\bibitem{na91} O. Nachtmann, Ann. Phys. 209 (1991) 436.

\bibitem{dfk}  H. G. Dosch, E. Ferreira, A. Kr\"amer, Phys. Rev.
  {\bf D50} (1994) 1992.

\bibitem{nico1}
  L. Lukaszuk, B. Nicolescu, Nuov. Cim. Lett. {\bf 8}, 405 (1973); 
  D. Joynson, E. Leader, C. Lopez, B. Nicolescu, Nuov. Cim.
  A {\bf 30}, 345  (1975)

\bibitem{odd}
  M. Rueter, H. G. Dosch, Phys. Lett. B 380, 177 (1996)

\bibitem{megg}
  E. Meggiolaro, Phys. Lett. B451, (1999) 414

\bibitem{donaru}
  M. Rueter , H.G. Dosch , O. Nachtmann, Phys. Rev. D59, 014018, 1999

\end{thebibliography}
\end{document}